# Fabrication and in vitro characterization of three-dimensional organic/inorganic scaffolds by robocasting

Julie Russias, Eduardo Saiz, Sylvain Deville, Karol Gryn, G. Liu, Ravi K. Nalla, Antoni P. Tomsia

Materials Sciences Division, Lawrence Berkeley National Laboratory, Berkeley, California 94720

**Abstract**

A key issue for the fabrication of scaffolds for tissue engineering is the development of processing techniques flexible enough to produce materials with a wide spectrum of solubility (bioresorption rates) and mechanical properties matching those of calcified tissues. These techniques must also have the capability of generating adequate porosity to further serve as a framework for cell penetration, new bone formation, and subsequent remodeling. In this study, we show how hybrid organic/inorganic scaffolds with controlled microstructures can be built using robotic assisted deposition at room temperature. Polylactide or polycaprolactone scaffolds with pore sizes ranging between 200–500 µm and hydroxyapatite contents up to 70 wt % were fabricated. Compressive tests revealed an anisotropic behavior of the scaffolds, strongly dependent on their chemical composition. The inclusion of an inorganic component increased their stiffness but they were not brittle and could be easily machined even for ceramic contents up to 70 wt %. The mechanical properties of hybrid scaffolds did not degrade significantly after 20 days in simulated body fluid. However, the stiffness of pure polylactide scaffolds increased drastically due to polymer densification. Scaffolds containing bioactive glasses were also printed. After 20 days in simulated body fluid they developed an apatite layer on their surface.

# INTRODUCTION

The demand for biomaterials to assist or replace organ functions and improve quality of life is rapidly increasing.[1] Traditional biomaterials for bone replacement are developed from materials designed originally for engineering applications that have serious shortcomings associated to the fact that their physical properties do not match those of the surrounding tissue and, unlike natural bone, cannot self-repair or adapt to changing physiological conditions. Thus, an ideal solution, and a scientific research challenge, is to develop bone-like biomaterials (or tissue engineering scaffolds) that will be treated by the host as normal tissue matrices and will integrate with bone tissue while they are actively resorbed or remodeled in a programmed way, with controlled osteogenic activity. This material will require an interconnected pore network with tailored surface chemistry for cell growth and penetration, and the transport of nutrients and metabolic waste. It should degrade at a controlled rate matching the tissue repair rates producing only metabolically acceptable substances and releasing drugs or stimulating the growth of new bone tissue at the fracture site by slowly releasing bone growth factors (e.g., bone morphogenic protein or transforming growth factor-$\beta$) throughout its degradation process. In addition, its mechanical properties should match those of the host tissues and the strength and stability of the material–tissue interface should be maintained while the material is resorbed or remodeled.

In recent years, there has been an increasing interest in the fabrication of porous three-dimensional structures with complex functionalities not only for use in tissue engineering but also for many other applications such energy generation, structural, electronics, etc. Techniques such as solvent casting, particulate leaching, gas foaming, etc.[2-5] have been traditionally used for synthesizing porous structures, but they suffer from numerous drawbacks including need for specialized tooling and molds, poor reproducibility, lack of proper control of porosity and interconnectivity, and, consequently, poor, unpredictable mechanical properties. This has been one of the motivations behind the development of new solid freeform fabrication techniques such as direct ink-jet printing, robotic assisted deposition or robocasting, and hot-melt printing, which usually involve "building" structures layer-by-layer by deposition of colloidal inks following a computer design; this can be achieved with great precision and reproducibility.[6-11]

One of the techniques that is gaining popularity during the last few years is robocasting[12, 13] where computer-controlled deposition of a thick slurry is performed to form three-dimensional structures layer-wise at room temperature. Until now, robocasting has been used to print ceramic inks where consolidation of the structures is achieved through a fluid through-gel transition during printing.[14] One of the great challenges is to develop the right inks or suspensions to print materials with a wider range of chemistries with precision and reproducibility. In this work, we demonstrate how this technique can be used to build three-dimensional organic/inorganic hybrid structures with controlled porosity, custom compositions, and appealing mechanical properties. The inorganic component of the materials is either hydroxyapatite (HA), an osteoconductive calcium phosphate closely related to the inorganic component of bone, or a bioactive silica-based glass. Since the formulation of the first bioactive glasses by Hench, they have been extensively used in the fabrication of biomaterials and composites due to their capacity to generate hydroxyapatite and form excellent bond with osseous tissue.[15] For the organic component we have chosen either polylactide (PLA) or polycaprolactone (PCL), two biocompatible and fully resorbable polymers with different stiffness, approved by the Food and Drug Administration (FDA) for medical applications. The work describes the development of hybrid inks for the fabrication of porous scaffolds by robotic assisted deposition and analyzes the physicochemical factors that control their final properties, in particular their in vitro evolution in simulated body fluid.

## MATERIALS AND METHODS

### Scaffolds fabrication

The inorganic component of the hybrid inks was either commercially available HA powders (Trans-tech Adamstown, MD, USA) with a particle size between 1 and 3 μm [Fig. 1(a)], Bioglass® with the composition 45S5 developed by Hench or a high-silica bioactive glass originally developed in our laboratory for the fabrication of coatings on metallic alloys (6P53B)16 (Table I). The glass powders have a wide particle size distribution with an average of ~13 ± 2 μm[16] [Fig. 1(b)]. Properties of the different materials used (density, Young's modulus, and compressive strength) are given in Table II.

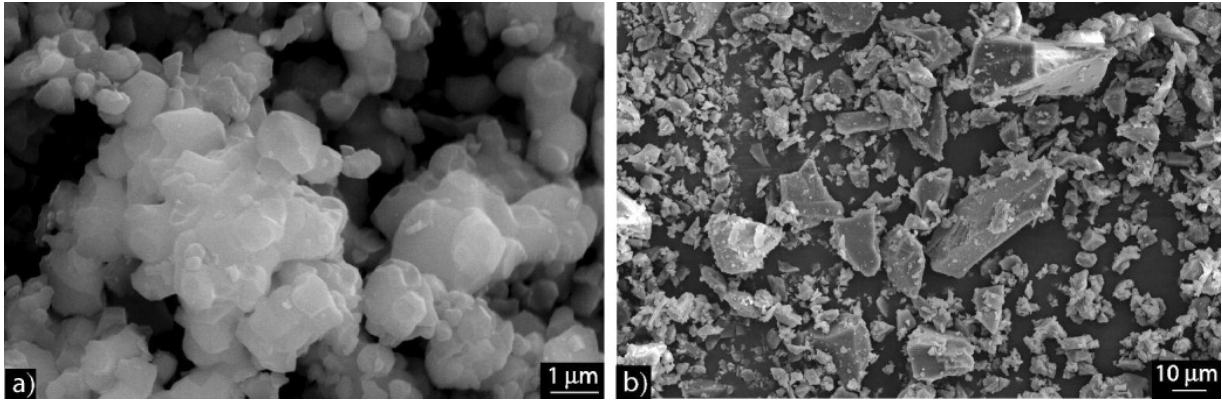

Figure 1: Scanning electron micrograph of (a) hydroxyapatite powders and (b) 6P53B glass powders used in this work.

Table I. 6P53B and Bioglass Compositions in Weight and mol %. Values in parentheses are in mol %.

|  | $SiO_2$ | $Na_2O$ | $K_2O$ | $CaO$ | $MgO$ | $P_2O_5$ |
|---|---|---|---|---|---|---|
| 6P53B | 52.7 (51.9) | 10.3 (9.8) | 2.8 (1.8) | 18.0 (19.0) | 10.2 (15.0) | 6.0 (2.5) |
| Bioglass | 45 (46.1) | 24.5 (24.3) | 0.0 (0.0) | 24.5 (26.9) | 0.0 (0.0) | 6.0 (2.6) |

Table II. Density, Young's Modulus, and Compressive Strength of the Raw Materials

|  | Density (g/cm³) | Young Modulus (GPa) | Compressive Strength (MPa) |
|---|---|---|---|
| PLA | 1.24 | 2.7[17] | 40–120 (pellet)[18] |
| PCL | 1.14 | 0.4[17] | – |
| Sintered dense HA | 3.16 | 35–120[19] | 120–900[19] |
| Bioglass® | 2.7 | 35 | ~500[18] |
| 6P53 B | 2.7[16] | 70[20] | ~70[21] |

Two different polymers were used in the inks: polylactide (PLA) (molecular weight = 92.1 kg/mol, 86.4% L isomer) and polycaprolactone (PCL) (Sigma Aldrich, Saint-Louis, MO, USA, molecular weight = 80 kg/mol). To prepare the inks, 5 g of polymer (PLA or PCL) were dissolved in 15 mL of methylene chloride (J.T. Baker, Phillipsburg, NJ, USA) at room temperature for 2 h using a magnetic stirrer. The required amount of HA or glass was added to the solution, together with 5 mL of denatured ethanol (water content <0.1%) to control the viscosity of the slurry and the evaporation kinetics and homogenized in a ball mill with alumina balls for 1 h. The final quality of the ink was assessed in terms of printability-measured as the minimum tip diameter suitable to extrude the ink without clogging and stability (i.e., shape retention capacity during drying and sintering) of the assembled structures.

Porous scaffolds were printed with a robocasting machine (3D inks, Stillwater, OK, USA) whose 3-axis motion was independently controlled by a custom-designed, computer aided direct-write program (Robocad 3.0, 3D inks, Stillwater, OK, USA). The deposition process was carried out in ambient atmosphere at room temperature. The three-dimensional periodic scaffolds (~15 × 15 mm2 in length and ~4 mm in height corresponding to 17 layers) consisted of a linear array of parallel rods in each layer aligned such that their orientation was orthogonal to the previous layer [Fig. 2(a)]. The center-to-center rod spacing was varied between 0.5 and 1 mm in order to change the porosity of the samples. Once a layer was printed, the nozzle was raised by a fixed height, which depends on the tip diameter, and another layer was deposited. The diameter of the printing nozzles was varied from 5 to 410 μm and the printing speed between 5 and 20 mm/s. During printing, the flow rate was adjusted to the nozzle diameter and the printing speed [Eq. (1)] in order to print a continuous line of uniform thickness:

$$V_p = \left(\frac{\varphi_t}{\varphi_s}\right)^2 V_w$$

where $V_p$ is the speed of the piston, $\phi_t$ is the diameter of the tip, $\phi_s$ is the diameter of the syringe, and $V_w$ is the printing speed. Samples were printed on glass slides and were easily removed after drying overnight at room temperature in air.

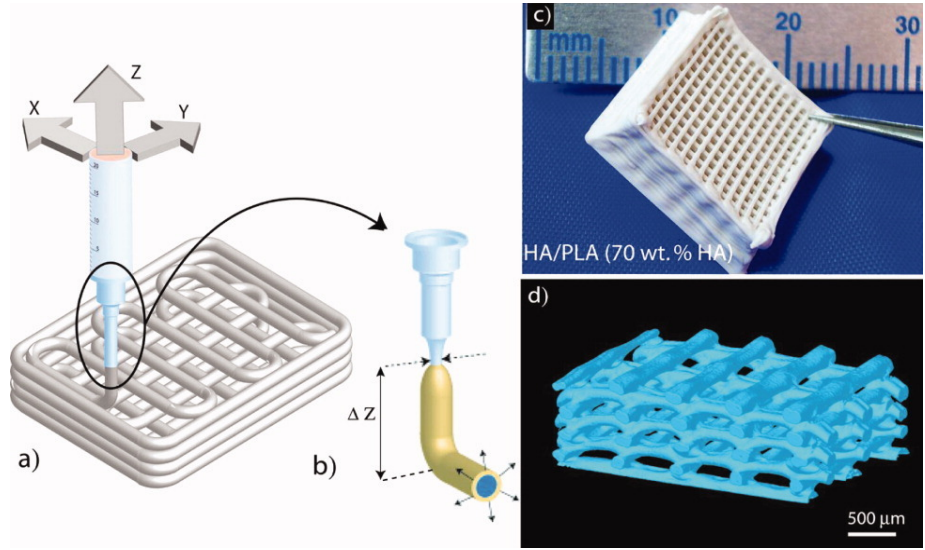

Figure 2: (a) Schematic representation of the robocasting process. The syringe displacement and the liquid flow through the nozzle are controlled by the computer. (b) During the printing, the inks swell after leaving the capillaries and the evaporation of the solvent creates a solid skin favoring the consolidation of the printed line. (c) Low magnification picture showing one of the scaffolds fabricates in this study (HA/PLA, 70 wt % HA scaffold with 17 layers). (d) Three-dimensional reconstructed image of a HA/PLA (70 wt % HA) grid obtained by synchrotron X-ray computed tomography. It can be observed that the material does not exhibit large defects

## Physical, microstructural, and mechanical characterization

The average porosity was calculated using the measured diameter of the printed rods and their spacing. The microstructure and composition of the printed parts were analyzed by X-ray diffraction (D500 diffractometer, Siemens AG, Munich, Germany), optical microscopy (Axiotech microscope, Carl Zeiss AG, Oberkochen, Germany), and environmental-scanning electron microscopy (ESEM: S-4300SE/N, Hitachi, USA) with associated energy dispersive spectroscopy (EDS). To obtain a three-dimensional perspective of the structure, particularly to reveal subsurface defects, synchrotron X-ray computed tomography was performed at the Advanced Light Source (ALS, Berkeley, CA). Imaging was achieved on representative structures with 26 keV monochromatic X-rays and a 6 μm voxel size (resolution). The tomography data were reconstructed into three-dimensional images by a Fourier-filtered back-projection algorithm as described in detail elsewhere.[22]

The Vickers micro hardness of the printed lines was measured by placing at least five indentations on 6 μm polished surfaces with a load of 30 g. Compression tests of strength, in the direction parallel and perpendicular to the printing plane, were carried out on a ELF® 3200 series voice-coil mechanical testing machine (EnduraTEC Inc., Minnetonka, MN) with a crosshead speed of 0.2 mm/min on pieces (4 × 3 × 3 mm3) cut with a blade from the printed scaffolds. At least four tests were performed for each composition and each direction.

The in vitro response of the scaffolds was studied by immersing the samples in 30 mL of simulated body fluid (Table III) at 37°C for 20 days. The solution was prepared by dissolving reagent-grade chemicals of NaCl, NaHCO3, KCl, K2HPO4, MgCl2.6H2O, CaCl2, and (CH2OH)3CNH2 into distilled water and buffered with HCl to pH 7.25 at 37°C.[23] The evolution of the pH in the solution was measured with a pH meter (model 3000, VWR Scientific). After 20 days of immersion, the samples were washed with large amounts of distilled water and air dried prior to microstructural and mechanical characterization.

Table III. Ion Concentrations of the SBF used in this Work and of Human Plasma

| Ion Concentration (mM) | $Na^+$ | $K^+$ | $Ca^{2+}$ | $Mg^{2+}$ | $Cl^-$ | $HCO_3^-$ | $HPO_4^{2-}$ | $SO_4^{2-}$ |
|---|---|---|---|---|---|---|---|---|
| SBF | 142.0 | 5.0 | 2.5 | 1.5 | 147.8 | 4.2 | 1.0 | 0.5 |
| Human plasma | 142.0 | 5.0 | 2.5 | 1.5 | 103.0 | 27.0 | 1.0 | 0.5 |

Thermal properties of PLA (in as received conditions, after 20 days at 37°C in air and after 20 days at 37°C in SBF) were determined using a differential scanning calorimeter (DSC). The measurements were run under Neon/Helium 50/50 volume mixture gas at a heating rate of 100°C/min using a Perkin Elmer Diamond DSC calibrated with indium. The typical sample weight was 1 mg. There was no thermal treatment of the samples before the first heating scan in order to preserve their history. The reported glass transition temperatures ($T_g$) were taken as the mid-point of the step transition based on first heating scans from 25 to 150°C.

## RESULTS AND DISCUSSION

### Physical and microstructural characterization of the printed samples

The process used in the preparation of the inks is based on the dissolution of the polymers in methylene chloride, like the standard procedure used for the preparation of biodegradable polymer and HA/polymer microspheres for drug delivery and composite fabrication.[24, 25] The two key differences are that no polyvinyl alcohol or any other difficult to eliminate and potentially toxic surfactants are used and that small quantities of a second organic phase (ethanol) are added in order to tailor the ink viscosity and drying rates to the printing conditions. During printing, solvent evaporation creates a solid skin on the extruded line immediately after it exits the tip [Fig. 2(b)]. This skin confers some degree of rigidity to the printed line and allows the fabrication of stable three-dimensional structures [Figs. 2(c,d)]. Consolidation of the lines is achieved by matching the printing speeds to the drying kinetics of the ink. For very slow printing speeds (typically less than 5 mm/s), the paste will dry up in the printing nozzle, thus clogging it; if the printing speed is too fast (faster than 20 mm/s), the line diameter is not homogeneous and the lines can be discontinuous.

The printing behavior of the inks also depends on their inorganic content. High particle loads can result in large viscosities and poor printability. However, by adjusting the amount of ethanol (the amount of ethanol vs. inorganics, i.e., ceramic powders, is ~40 wt %) it is possible to prepare hybrid scaffolds with inorganic contents as large as 70 wt %, very close to the mineral content of cortical bone. This is very important since the ceramic content influences the stiffness, a key mechanical parameter that should be matched with the host tissue. Typical required Young's modulus (E) varies between 0.4 and 350 MPa for soft tissue and cartilage and up to 10–1500 MPa for hard tissue.6 Particular attention should be paid to the control of the evaporation of the organic component during ink preparation. Because of the use of methylene chloride as a solvent, drying is quite rapid and the ink properties need to be maintained in the range adequate for printing by hermetically sealing the containers. However, because the main mechanism of consolidation is drying, control of the rheological properties does not need to be as strict as for ceramic inks.[13] Furthermore, during robocasting of ceramic parts different strategies (such as printing in oil baths) are used to avoid drying and the associated dimensional changes and stresses during printing.[26] In the process described in this work, the viscoplastic behavior of the parts allows them to sustain the stresses generated during simultaneous printing and drying. This viscoplastic nature permits further stretching of the printed line to form thin polymer and polymer/ceramic fibers and ribbons (down to 1 μm in diameter). Figure 3 shows very thin polymer and hybrid threads fabricated in this manner. It can be clearly observed that the minimum achievable diameter of the thread is controlled by the size of ceramic particles. Biodegradable fibers can be used in drug delivery applications or in the fabrication of meshes for implants and scaffolds.[27]

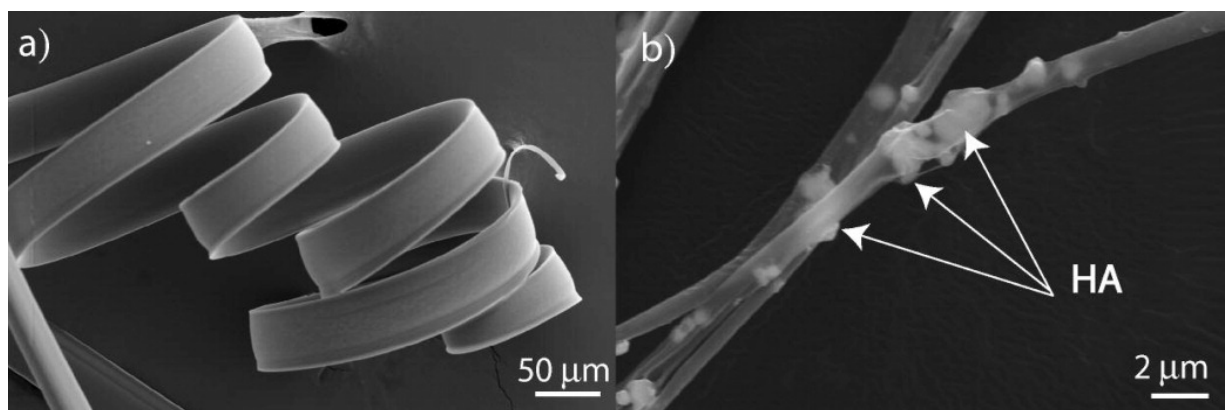

Figure 3: (a) Pure PLA flexible ribbon and (b) thin HA/PLA (30/70 wt %) thread extruded after printing. Hydroxyapatite grains are embedded in the polymer fiber and determine its minimum diameter.

It was not possible to successfully prepare an ink containing the Bioglass composition developed by Hench. This is attributed to the highly hygroscopic nature of this glass (related to its high bioactivity) that promotes fast reactions with small amounts of water in the environment or in the ink. This fact also hampers the use of Bioglass in the preparation of suspensions to coat metallic alloys by enameling[28] and motivated the development of a family of high-silica bioactive glasses like 6P53B that are much less hygroscopic and more amenable to the preparation of suspensions. Although several techniques have been proposed to prepare Bioglass/polymer foams and porous materials[18, 29, 30] by using the 6P53B composition, it was possible to prepare hybrid inks optimized for rapid prototyping.

Partial wetting of the ink on the printed material increases the contact area and provides better bonding between lines [Fig. 4(a)], enhancing the stability of the structure. In the hybrid scaffolds the ceramic phase is homogeneously distributed in the polymer [Fig. 4(b)] and there are no visible surface defects or microporosity. By changing the center-to-center rod spacing (0.5 and 1 mm), samples with 2 different porosities are processed. A center-to-center rod spacing of 0.5 mm corresponds to a porosity of 55% and a pore size of about 200 μm [Fig. 4(c)], while a center-to-center rod spacing of 1 mm gives a porosity of 75% and a pore size of about 500 μm [Fig. 4(d)].

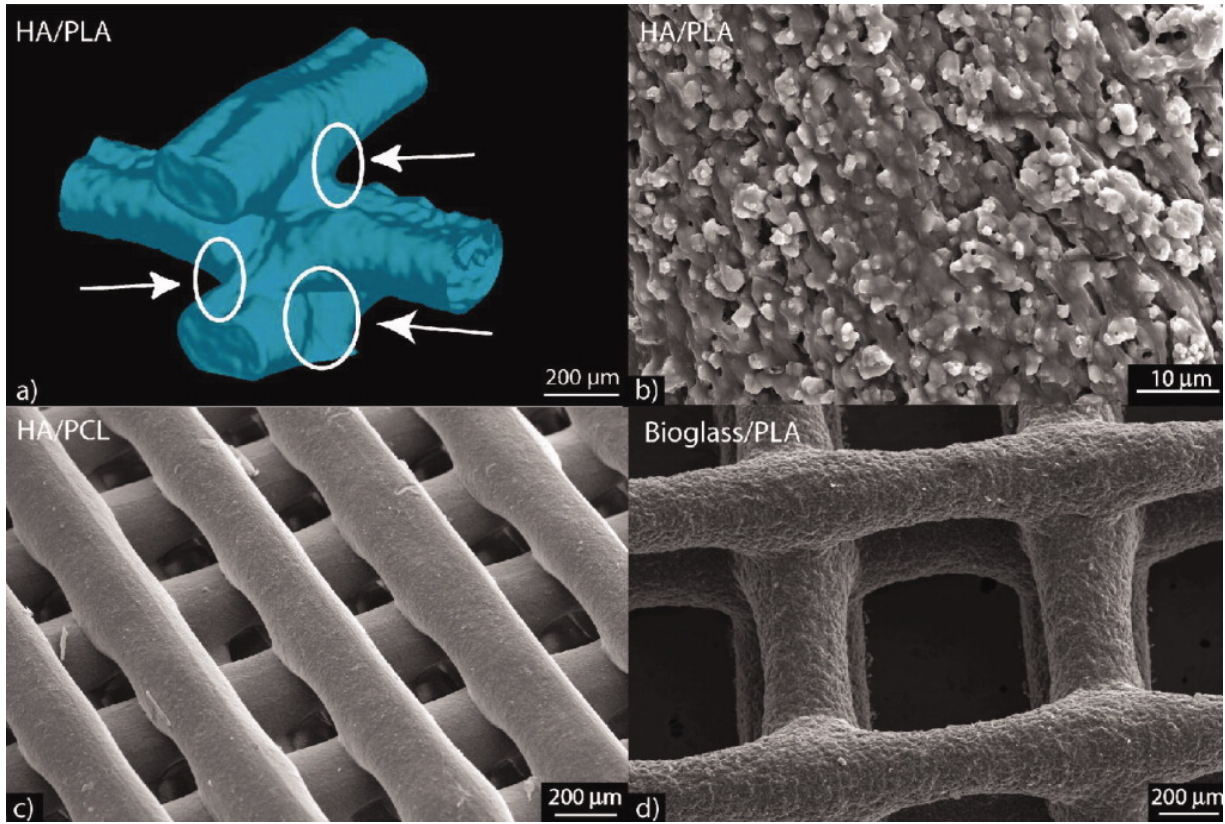

Figure 4: Scanning electron micrographs showing different aspects of the microstructure. All the scaffolds showed in this picture have been printed using a nozzle with an internal diameter of 410 μm. (a) Three-dimensional reconstructed image of the junction between HA/PLA (70 wt % HA) printed lines obtained by synchrotron X-ray computed tomography. It can be observed that the ink wets partially the printed lines providing additional support and leading to good bonding between lines (white circles). (b) Close up of an HA/PLA (70 wt % HA) printed line showing the homogeneous distribution of the HA particles (white) in the polymer. (c) HA/PCL (70 wt % HA) scaffold (center-to-center rod spacing ~0.5 mm, corresponding to a porosity of ~55%); the pore size and thickness of the printed lines is very homogeneous. (d) 6P53B glass/PLA (70 wt % 6P53B glass) scaffold (center-to-center rod spacing ~1 mm, corresponding to a porosity of ~75%).

The final line thickness depends on several factors: nozzle diameter, printing speed, printing height [$\Delta Z$ in Fig. 2(b)], and drying shrinkage (typically ~25% in volume). After leaving the nozzle, the ink was observed to swell, with the degree of swelling depending on the printing speed (flow rate) and the nozzle diameter. As shown in Figure 5, for a 410-μm tip, the swelling and the printing speed influence are minimal and the drying shrinkage is important since the final average line diameter is around 200 µm. For a very thin capillary of 5 µm, there is a significant swelling that can be controlled by increasing the printing speed and the diameter of the printed lines can be reduced from 400 to 100 µm, without affecting the pore size, by changing the printing speed from 5 to 20 mm/s.

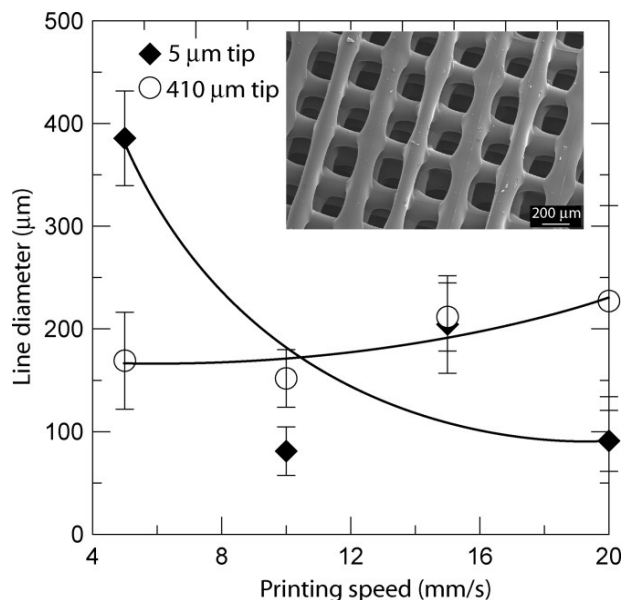

Figure 5: Evolution of the final line diameter versus the printing speed for the printing of a pure PLA ink through nozzles with two internal diameters. The insert shows a scanning electron micrograph of a PLA grid printed with a 5-µm tip at a printing speed of 15 mm/s.

In a further step, the scaffolds can be infiltrated to make dense samples with controlled composition and phase distributions. By using phases with different degradation rates porosity can be created "in situ" providing interesting possibilities for the control of the bioresoption and the mechanical behavior in vivo. For example, we have been able to successfully infiltrate a PLA/HA (70 wt % of HA) scaffold with a PCL/HA slurry (70 wt % of HA, with a slurry prepared exactly in the same conditions than the inks for printing) under vacuum (150 mbar) without leaving residual porosity (Fig. 6). The systematic analysis of the microstructure and properties of these materials will be the subject of a future work.

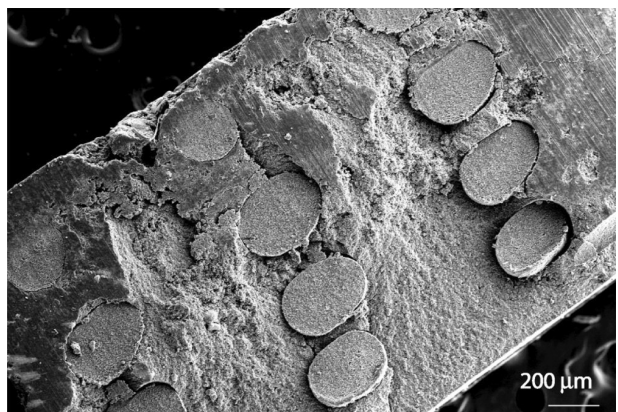

Figure 6: Scanning electron micrograph of a PLA/HA (70 wt % HA) scaffold cross section infiltrated with PCL/HA (70 wt % HA). No residual porosity can be observed after the infiltration process.

**In vitro behavior in simulated body fluid**

In vitro tests in cell-free solutions with ionic concentrations similar to those of body fluids allows analysis of the chemical and microstructural evolution of the materials under conditions that simulate their biological interactions with the body and provide fundamental data to predict and understand their in vivo behavior and long term stability.[23] After 20 days in simulated body fluid microporosity develops on the surface of pure PLA scaffolds (pore size from 0.5 to 1 µm) (Fig. 7). To analyze the possible changes in the polymer structure induced by the in vitro treatment at 37°C, DSC analysis of the "as received" PLA and after 20 days at 37°C in air and in SBF are performed (Fig. 8). The DSC analysis confirms that, as expected, the as received PLA is amorphous since it contains 14% of D isomer. However, its glass transition temperature increases from 33.0°C for the as received materials to 49.1°C after 20 days at 37°C or to 60.8°C after 20 days at 37°C in SBF. These results indicate that the structure of the polymer is changing during the in vitro tests: the polymer chains are organizing and the polymer is getting denser. This temperature and humidity are two factors responsible for the densification of the PLA. No diffusion of SBF ions into the polymer is observed by EDS analyses. The volume change associated with the densification might be responsible for the formation of the micropores. The addition of HA to the PLA reduces significantly the shrinkage of the material and micron size pores do not form on HA/PLA scaffolds even after 20 days in SBF. Degradation of the PCL-based scaffolds (with or without hydroxyapatite) results in the formation of a network of polymer fibers between the printed lines (Fig. 9) while their surface becomes rougher.

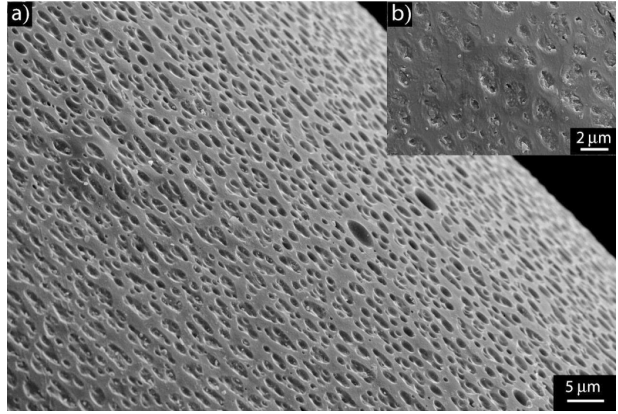

Figure 7: (a) Scanning electron micrographs of a PLA scaffold after 20 days in SBF. Note the presence of micron size pores homogeneously dispersed on the polymer surface. (b) The inset shows these pores at a larger magnification.

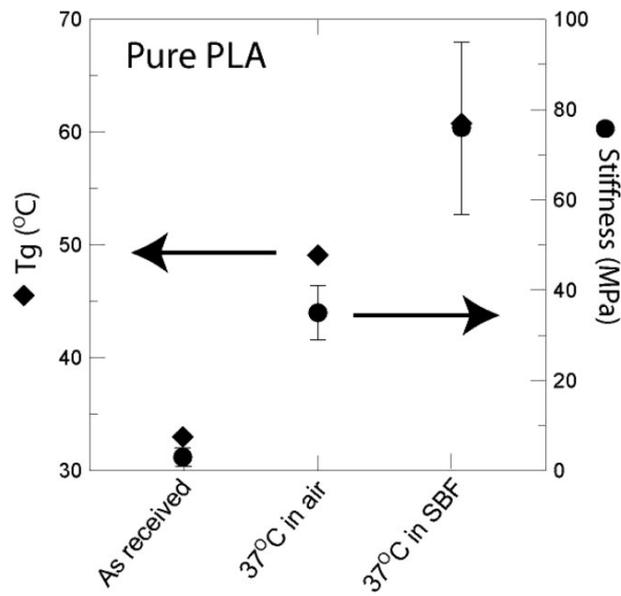

Figure 8: Evolution of the glass transition temperature and stiffness up to the elastic limit for PLA scaffolds in an as received condition, after 20 days at 37°C in air and after 20 days at 37°C in simulated body fluid.

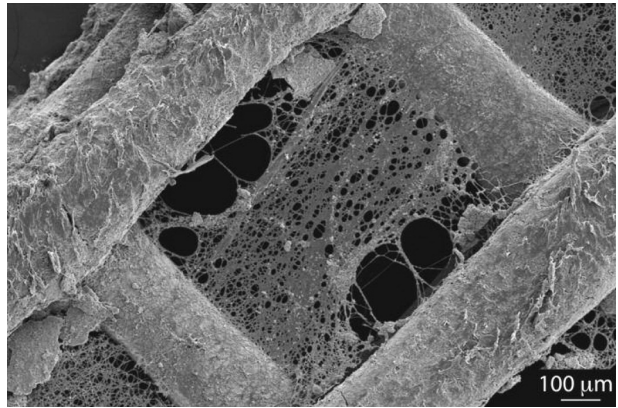

Figure 9: Scanning electron micrograph of a PCL/HA (70 wt % HA) scaffold after 20 days in SBF. Polymer filaments are bridging the lines. The degradation of the scaffold surface is clearly visible.

The formation of acidic products during the hydrolytic degradation of resorbable polymer materials has been reported to cause adverse body reactions and additions of various calcium phosphates to the polymer have been proposed as a mean to buffer the release of acidic products and avoid these reactions.[31-34] Figure 10 shows the pH evolution of the simulated body fluid solution in which the PLA-based samples were immersed. The behavior is similar for PCL based scaffolds. The pH remains stable, around 7.25, for pure polymer and HA/polymer composites. The presence of glass particles has two effects: the pH of the solution increases to reach 7.8 after 20 days of immersion for scaffolds containing 6P53B glass and apatite crystals precipitate on the scaffold surface (Fig. 11). This is due to a rapid ion exchange of Na+ from the glass with H+ and H3O+ followed by a polycondensation reaction of surface silanols to create a high-surface area silica gel. This gel can provide a large number of sites for heterogeneous nucleation and crystallization of a biologically reactive hydroxy-carbonate apatite (HCA) layer equivalent to the inorganic mineral phase of bone.[15] The ion exchange from the bioactive glass can buffer the acidic products resulting from polymer degradation and promote apatite precipitation. It has been proposed that this growing apatite layer favors the bonding to bone of bioactive glasses and has a significant impact on the activity of osteogenic cells.[35]

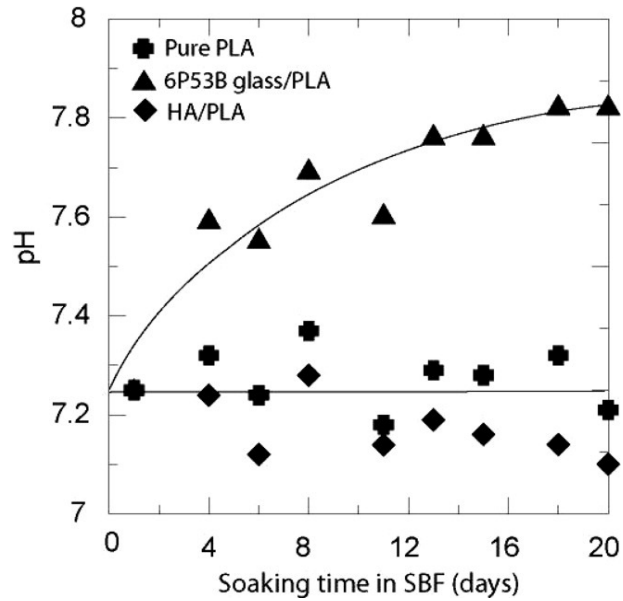

Figure 10: pH evolution of the solution during 20 days of immersion in SBF for PLA, PLA/HA, and PLA/6P53B glass materials. Note the increase in pH for the 6P53B glass/PLA scaffold.

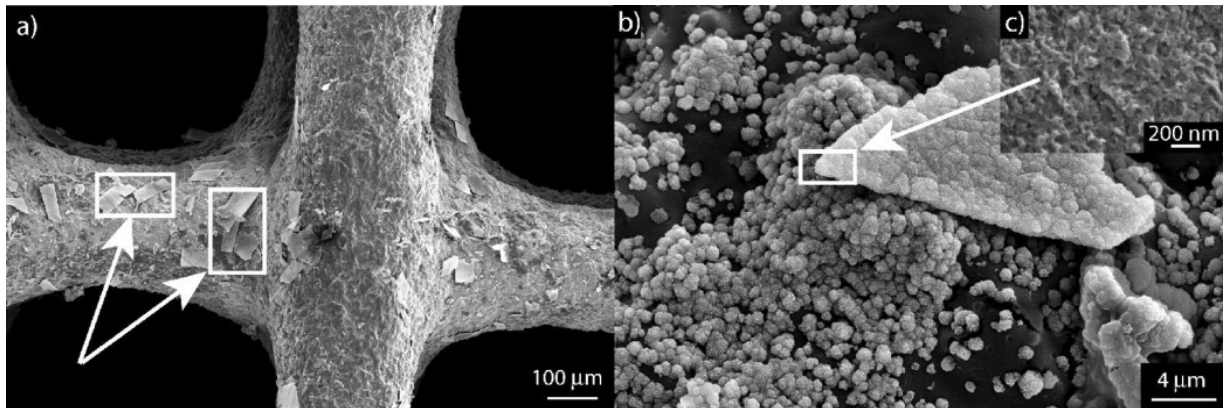

Figure 11: Scanning electron micrographs of the 6P53B glass/PLA scaffold after 20 days of immersion in SBF. (a) Apatite crystals have precipitated on the scaffold surface. (b) The apatite layers consist of small nano-crystals (inferior to 50 nm) that cover progressively the scaffold.

## Mechanical characterization

### Microhardness

As expected from the intrinsic properties of the polymers (Table II) the printed lines of the PLA-based scaffolds are much harder than the equivalent PCL composites (Fig. 12). The hardness of the glass is 6.2 GPa.[16] There is a wide spread in the hardness data reported for sintered HA but it seems to range between 3.5 and 6.5 GPa for a fully dense ceramic.[36-38] The addition of glass increases the hardness but there is a larger dispersion in the measurements probably due to the larger grain size and the larger dispersion of the grain size distribution. After immersion in simulated body fluid the hardness of pure PLA and PLA/HA materials increase slightly. This might be due to the observed densification of the polylactide during the in vitro tests.

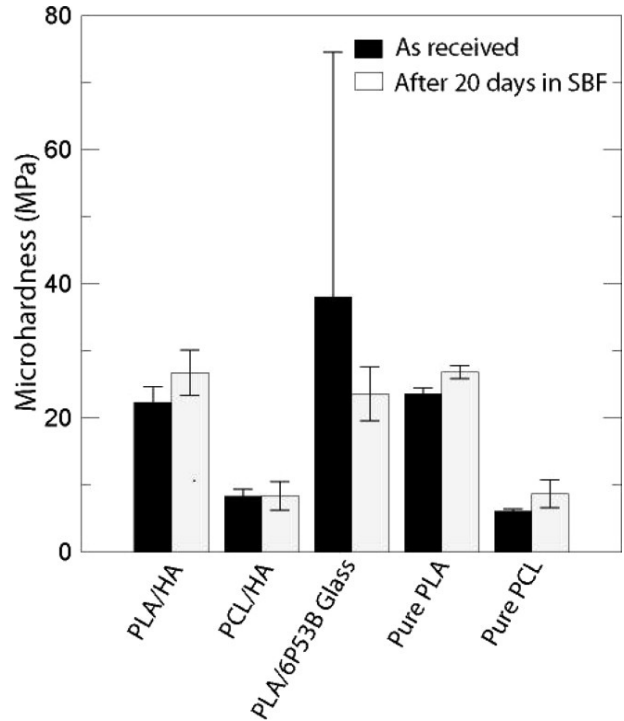

Figure 12: Micro hardness of the printed lines samples in as received conditions and after 20 days in a SBF. Organic/inorganic samples have 70 wt % of inorganic phase.

## Compression strengths

Compression tests are performed in two directions (Fig. 13), perpendicular (direction 1) and parallel (direction 2) to the printing plane. During the tests, the scaffolds do not fail in a brittle manner and show an elasto-plastic response with large plastic "yielding". The insets on Figure 13 show SEM pictures of the samples in both directions after compression. Only PLA/HA samples tested in the direction 2 show a maximum in the stress vs. strain curve at around 7–10 MPa. This is probably due to the fact that they are harder and stiffer and the printed lines buckle instead of being continuously deformed like in the other scaffolds (Fig. 13).

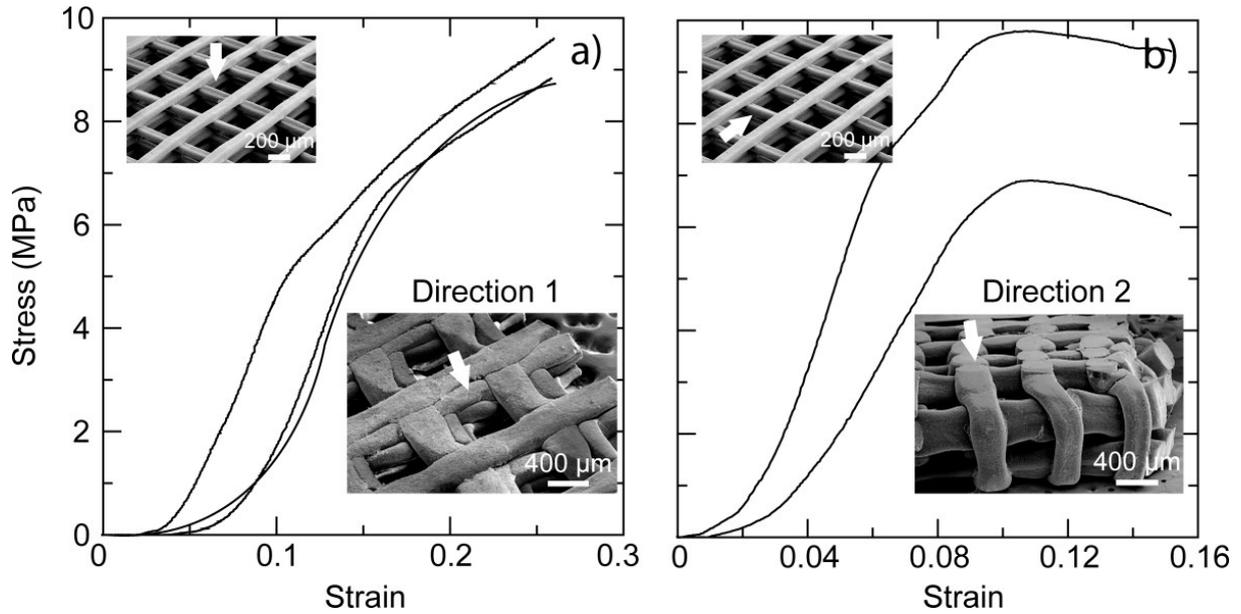

Figure 13: Strain–stress curves obtained from compression tests made on different grids of HA/PLA (70 wt % HA) samples with 55% of porosity in the two directions (a) perpendicular (direction 1) and (b) parallel (direction 2) to the printing plane. The insets show some SEM images of the samples after compression tests in direction 1 and 2. The stress–strain curves in each direction are very similar. The samples do not fail in a brittle manner and the behavior is anisotropic. Buckling of the printed lines of the PLA/HA scaffolds can be clearly observed in the inset [Fig. 13(b)].

The lines do not debond during testing in any direction, indicating excellent adhesion between the printed rods. The stress–strain curves for samples with the same composition and microsctructure are very similar (Fig. 13), allowing the comparison of the mechanical response of different materials. Considering the samples' geometry and due to their large deformation, it is difficult to calculate an absolute value of the Young's modulus from the compression tests. However, the slope of the compression curves (calculated using a linear fit of the data up to the elastic limit) can provide several trends.

1. The porosity (in the range used in this work, between 55 and 75 vol %) does not affect significantly the mechanical response of the samples.
2. The mechanical response of the scaffolds is clearly anisotropic (Fig. 13).
3. The mechanical properties can be easily adjusted by controlling the composition of the material: type of polymer and inorganic content. Addition of up to 70 wt % of hydroxyapatite increases the stiffness up to two orders of magnitude [Fig. 14(a)] and PLA-based scaffolds are much stiffer than those containing PCL [Fig. 14(b)]. For an HA/polymer scaffolds with a porosity of 55% (pore size ~200 × 200 μm2) and an HA content of 70 wt %, the values of the slopes in the direction perpendicular to the printing plane varies between 84 ± 9 MPa for a PLA-based materials and 24 ± 5 MPa for PCL-based materials. In the parallel direction, the slopes vary between 150 ± 40 and 110 ± 20 MPa for PLA and PCL-based scaffolds, respectively. This is expected since the elastic modulus of dense PLA and PCL are respectively 2.7 and 0.4 GPa (Table II).

4. After twenty days in SBF, pure PLA scaffolds become much stiffer and the slope of the strain–stress curve up to the elastic limit increases from 3 ± 2 to 76 ± 19 MPa [Fig. 15(a)]. Like the parallel increase in microhardness, this is related to the observed densification of the polymer as a result of the in vitro treatment. The PLA scaffolds also become stiffer after 20 days in air at 37°C. The slope's average value of their strain–stress curve up to the elastic limit is 35 ± 6 MPa. As it is confirmed by the DSC analysis, the long treatment at 37°C promotes densification and this process is enhanced in SBF (Fig. 8). The moisture plasticizes the polymer chains favoring their rearrangement and enhancing densification. This result suggests that thermal treatments can be used to further manipulate the mechanical response of the material.

5. The compressive behavior of the PLA or PCL-based hybrid scaffolds containing 70 wt % of hydroxyapatite do not change significantly after 20 days in simulated body fluid. The elastic modulus of HA is 10–100 times larger than the one of the polymer (Table II) and the main contribution to the Young's modulus of the composites comes from the inorganic phase that does not undergo any visible degradation during in vitro testing in SBF. However, the bioactive glasses react with simulated body fluid and there is an appreciable decrease of the stiffness in the scaffolds containing 6P53B after 20 days in SBF [Fig. 15(b)].

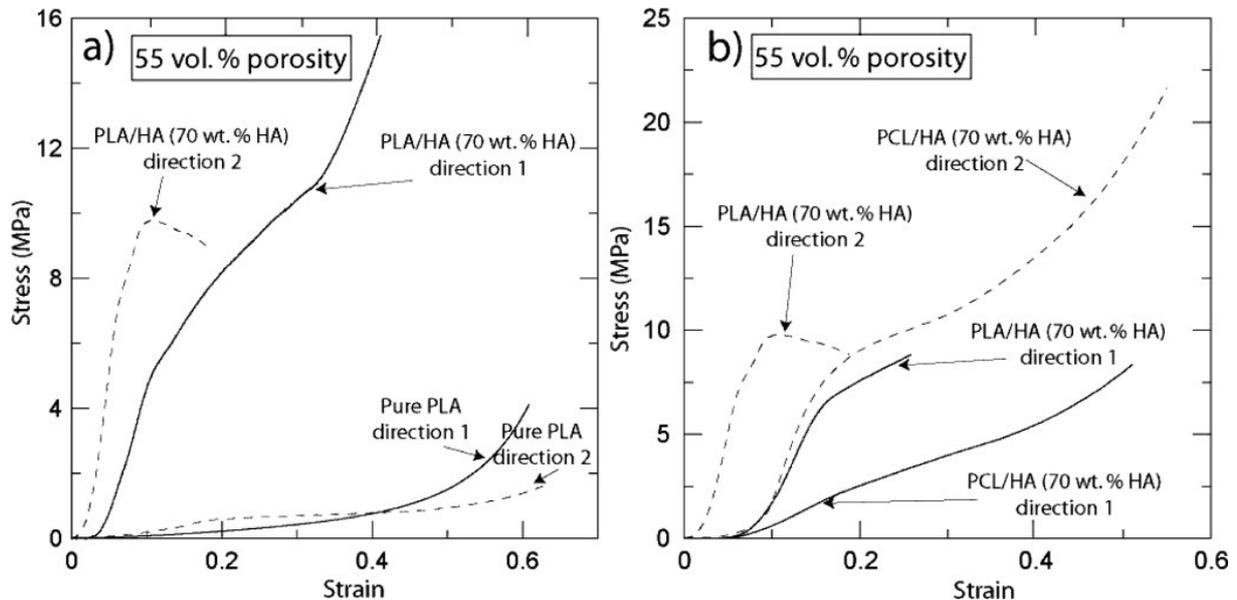

Figure 14: (a) Influence of the ceramic content on the compressive behavior of the scaffolds. The strain–stress curves are obtained from compression tests made on PLA and HA/PLA samples in two directions. Addition of hydroxyapatite increases significantly the stiffness of the material. (b) Influence of the organic phase. Samples containing PLA are stiffer.

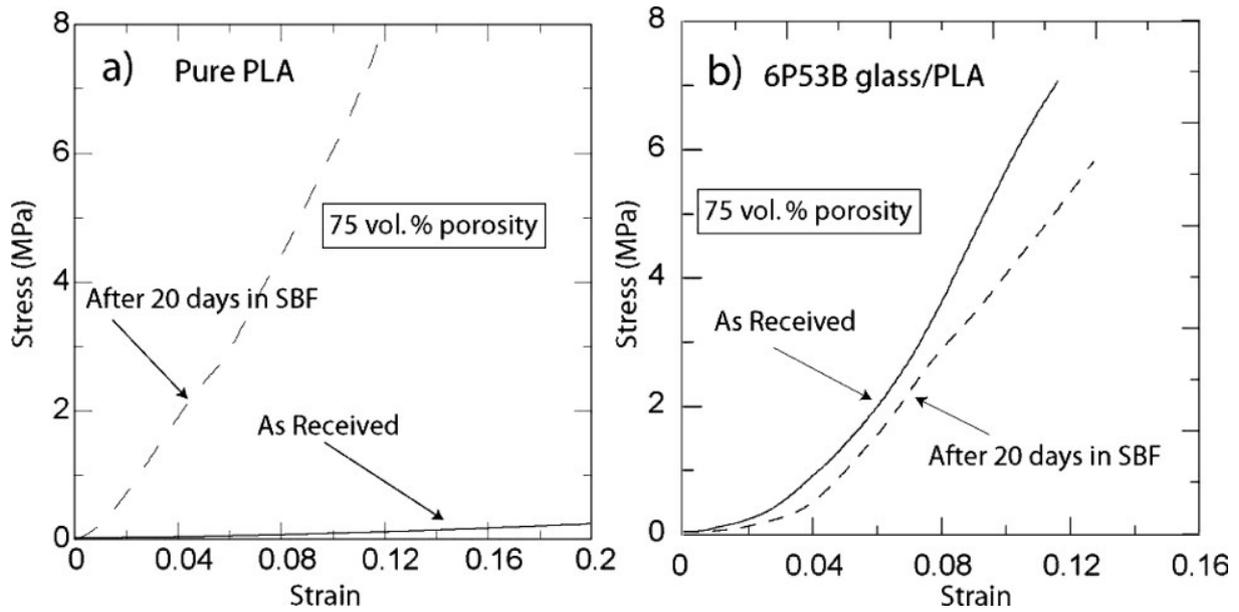

Figure 15: (a) Strain–stress curves obtained from compression tests made on pure PLA samples in the direction 1. The sample has become stiffer after in vitro testing in SBF. (b) Strain–stress curves obtained from compression tests made on 6P53B glass/PLA (70 wt % 6P53B glass) samples in the direction 1. Before immersion, 6P53B glass/PLA composites have a slope's average value of their strain–stress curve up to the elastic limit of 105 ± 18 MPa. This value is comparable with the one of the as received HA/PLA in direction 1. After 20 days in a physiological environment, the slope's average value of the composites with bioactive glass decreases to 62 ± 10 MPa (~40% reduction) due to a partial dissolution of the glass in the solution before the formation of apatite crystals at the surface of the scaffold.

## SUMMARY

This work demonstrates how robotic assisted deposition can be successfully used for the fabrication of porous hybrid organic/inorganic materials of various chemical compositions with well controlled architecture and porosity. The technique is versatile enough to allow the combination of a wide range of materials, including bioactive glasses whose addition can be used to buffer the possible formation of acidic degradation product coming from the hydrolytic degradation of the polymer and to promote apatite formation. In addition, by selecting the adequate organic component and manipulating the organic/inorganic ratio of the scaffolds, we can control their stiffness and fabricate materials much more rigid than porous polymers while avoiding the brittleness of ceramic parts. Moreover, multi-component materials can be fabricated through the simultaneous use of multiple nozzles and because the processing is performed at room temperature, in situ seeding with cells and addition of drugs or growth factors to the organic component is easily achievable. These characteristics suggest that robotic assisted deposition can be a fast and economical alternative for the fabrication of "on demand" scaffolds for biomedical applications.

## Acknowledgements

We acknowledge the support of the dedicated tomography beamline (BL 8.3.2) at the Advanced Light Source (ALS).

## References

1. Service RF. Tissue engineers build new bone. *Science* 2000; 289: 1498–1500.

2. Ambrosio AMA, Sahota JS, Khan Y, Laurencin CT. A novel amorphous calcium phosphate polymer ceramic for bone repair. Part 1: Synthesis and characterization. *J Biomed Mater Res* 2001; 58: 295–301.

3. Mathieu LM, Mueller TL, Bourban PE, Pioletti DP, Müller R, Månson JAE. Architecture and properties of anisotropic polymer composite scaffolds for bone tissue engineering. *Biomaterials* 2006; 27: 905–916.

4. Kothapalli CR, Shaw MT, Wei M. Biodegradable HA-PLA 3D porous scaffolds: Effect of nano-sized filler content on scaffold properties. *Acta Biomaterialia* 2005; 1: 653–662.

5. Cheng W, Li HY, Chang J. Fabrication and characterization of β-dicalcium silicate/poly(D,L-lactic acid) composite scaffolds. *Mater Lett* 2005; 59: 2214–2218.

6. Hollister SJ. Porous scaffold design for tissue engineering. *Nat Mater* 2005; 4: 518–524.

7. Roy TD, et al. Performance of degradable composite bone repair products made via three-dimensional fabrication techniques. *J Biomed Mater Res A* 2003; 66: 283–291.

8. Xiong Z, Yan YN, Wang SG, Zhang RJ, Zhang C. Fabrication of porous scaffolds for bone tissue engineering via low-temperature deposition. *Scripta Mater* 2002; 46: 771–776.

9. Kalita SJ, Bose S, Hosick HL, Bandyopadhyay A. Development of controlled porosity polymer–ceramic composite scaffolds via fused deposition modeling. *Mater Sci Eng C: Biol Sci* 2003; 23: 611–620.

10. Chu TMG, Orton DG, Hollister SJ, Feinberg SE, Halloran JW. Mechanical and in vivo performance of hydroxyapatite implants with controlled architectures. *Biomaterials* 2002; 23: 1283–1293.

11. Grida I, Evans JRG. Extrusion freeforming of ceramics through fine nozzles. *J Eur Ceram Soc* 2003; 23: 629–635.

12. Cesarano IJ, Calvert P. Freeforming objects with low-binder slurry. United States Patent 6027326 US Patent Issued on February 22, 2000.

13. Smay JE, Cesarano J, Lewis JA. Colloidal inks for directed assembly of 3-D periodic structures. *Langmuir* 2002; 18: 5429–5437.

14. Michna S, Wu W, Lewis JA. Concentrated hydroxyapatite inks for direct-write assembly of 3-D periodic scaffolds. *Biomaterials* 2005; 26: 5632–5639.

15. Hench LL, Polak JM. Third-generation biomedical materials. *Science* 2002; 295: 1014–1017.

16. Lopez-Esteban S, et al. Bioactive glass coatings for orthopedic metallic implants. *J Eur Ceram Soc* 2003; 23: 2921–2930.

17. Murugan R, Ramakrishna S. Development of nanocomposites for bone grafting. *Compos Sci Technol* 2005; 65: 2385–2406.

18. Rezwan K, Chen QZ, Blaker JJ, Boccaccini AR. Biodegradable and bioactive porous polymer/inorganic composite scaffolds for bone tissue engineering. *Biomaterials* 2006; 27: 3413–3431.

19. Suchanek W, Yoshimura M. Processing and properties of hydroxyapatite-based biomaterials for use as hard tissue replacement implants. *J Mater Res* 1998; 13: 94–117.

20. Pavón J, Jiménez-Piqué E, Anglada M, Saiz E, Tomsia AP. Monotonic and cyclic Hertzian fracture of a glass coating on titanium-based implants. *Acta Mater* 2006; 54: 3593–3594.

21. Goller G, Demirkiran H, Oktar FN, Demirkesen E. Processing and characterization of bioglass reinforced hydroxyapatite composites. *Ceram Int* 2003; 29: 721–724.

22. Kinney JH, Nichols MC. X-Ray tomographic microscopy (XTM) using synchrotron radiation. *Annu Rev Mater Sci* 1992; 22: 121–152.

23. Kokubo T, Kushitani H, Sakka S, Kitsugi, T, Yamamuro T. Solutions able to reproduce in vivo surface–structure change in bioactive glass–ceramic A-W. *J Biomed Mater Res* 1990; 24: 721–734.

24. ODonnell PB, McGinity JW. Preparation of microspheres by the solvent evaporation technique. *Adv Drug Deliver Rev* 1997; 28: 25–42.

25. Russias J, Saiz E, Nalla RK, Tomsia AP. Microspheres as building blocks for hydroxyapatite/polylactide biodegradable composites. *J Mater Sci* 2006; 41: 5127–5133.

26. Miranda P, Saiz E, Gryn K, Tomsia AP. Sintering and robocasting of β-tricalcium phosphate scaffolds for orthopedic applications. *Acta Biomaterialia* 2006; 2: 457–466.

27. Luong-Van E, et al. Controlled release of heparin from poly(E-caprolacptone) electrospun fibers. *Biomaterials* 2006; 27: 2042–2050.

28. Pazo A, Saiz E, Tomsia AP. Silicate glass coatings on Ti-based implants. *Acta Mater* 1998; 46: 2551–2558.

29. Kazarian SG, Chan KLA, Maquet V, Boccaccini AR. Characterisation of bioactive and resorbable polylactide/Bioglass(R)) composites by FTIR spectroscopic imaging. *Biomaterials* 2004; 25: 3931–3938.

30. Chen QZ, Boccaccini AR. Poly(D,L-lactic acid) coated 45S5 bioglass (R)-based scaffolds: Processing and characterization. *J Biomed Mater Res A* 2006; 77: 445–457.

31. Hutmacher DW. Scaffolds in tissue engineering bone and cartilage. *Biomaterials* 2000; 21: 2529–2543.

32. Bergsma JE, Debruijn WC, Rozema FR. Late degradation tissue-response to poly(lactide) bone plates and screws. *Biomaterials* 1995; 16: 25–31.

33. VanderGiessen WJ, Lincoff AM, Schwartz RS. Marked inflammatory sequelae to implantation of biodegradable and nonbiodegradable polymers in porcine coronary arteries. *Circulation* 1996; 94: 1690–1697.

34. Trantolo DJ, Gresser JD, Wise DL, Lewandrowski K. Buffered biodegradable internal fixation devices. In: Donald L. Wise, editor. Biomaterials and Bioengineering Handbook. New York: Marcel Dekker; 2000.

35. Hench LL. Biomaterials: A forecast for the future. *Biomaterials* 1998; 19: 1419–1423.

36. Hoepfner TP, Case ED. The influence of the microstructure on the hardness of sintered hydroxyapatite. *Ceram Int* 2003; 29: 699–706.

37. Kumar RR, Wang M. Modulus and hardness evaluations of sintered bioceramic powders and functionally graded bioactive composites by nano-indentation technique. *Mat Sci Eng A* 2002; 338: 230–236.

38. Thangamani N, Chinnakali K, Gnanam FD. The effect of powder processing on densification, microstructure and mechanical properties of hydroxyapatite. *Ceram Int* 2002; 28: 355–362.